\documentclass[journal]{IEEEtran}
\usepackage{cite}
\usepackage{amsmath,amssymb,amsfonts}
\usepackage{subcaption}
\usepackage{algorithm}
\usepackage{algpseudocode}
\usepackage{graphicx}
\usepackage{textcomp}
\def\BibTeX{{\rm B\kern-.05em{\sc i\kern-.025em b}\kern-.08em
    T\kern-.1667em\lower.7ex\hbox{E}\kern-.125emX}}
\begin{document}
\title{Patlak Parametric Image Estimation from Dynamic PET Using Diffusion Model Prior}
\author{Ziqian Huang, Boxiao Yu, Siqi Li, Savas Ozdemir, Sangjin Bae, Jae Sung Lee, Guobao Wang, Kuang Gong
\thanks{This work was supported by NIH grants R01EB034692 and R01AG078250. (Corresponding author: Kuang Gong) }
\thanks{Z. Huang, B. Yu, and K. Gong are with J. Crayton Pruitt Family Department of Biomedical Engineering, University of Florida, Gainesville, FL 32601 USA (e-mail:ziqian.huang@ufl.edu, boxiao.yu@ufl.edu, kgong@bme.ufl.edu). }
\thanks{S. Ozdemir is with the Department of Radiology, University of Florida, Jacksonville, FL 32209 USA (e-mail:savas.ozdemir@jax.ufl.edu). }
\thanks{S. Bae is with the Interdisciplinary Program in Bioengineering, Seoul National University, Seoul, 08826, South Korea (e-mail:rubp3pg@snu.ac.kr).}
\thanks{J. Lee is with the Department of Nuclear Medicine, Seoul National University College of Medicine, Seoul 03080, South Korea (e-mail:jaes@snu.ac.kr). }
\thanks{S. Li and G. Wang are with the Department of Radiology, University of California Davis Health, Sacramento, 95817, CA, USA (e-mail:sqlli@ucdavis.edu, gbwang@ucdavis.edu). }}

\maketitle

\begin{abstract}
Dynamic PET enables the quantitative estimation of physiology-related parameters and is widely utilized in research and increasingly adopted in clinical settings. Parametric imaging in dynamic PET requires kinetic modeling to estimate voxel-wise physiological parameters based on specific kinetic models. However, parametric images estimated through kinetic model fitting often suffer from low image quality due to the inherently ill-posed nature of the fitting process and the limited counts resulting from non-continuous data acquisition across multiple bed positions in whole-body PET. In this work, we proposed a diffusion model-based kinetic modeling framework for parametric image estimation, using the Patlak model as an example. The score function of the diffusion model was pre-trained on static total-body PET images and served as a prior for both Patlak slope and intercept images by leveraging their patch-wise similarity. During inference, the kinetic model was incorporated as a data-consistency constraint to guide the parametric image estimation. The proposed framework was evaluated on total-body dynamic PET datasets with different dose levels, demonstrating the feasibility and promising performance of the proposed framework in improving parametric image quality.
\end{abstract}


\section{Introduction}
\label{sec:introduction}
\IEEEPARstart{P}{ositron} emission tomography (PET) is a functional imaging modality that employs radiotracers to visualize and quantify physiological processes. Compared to static PET, dynamic PET provides additional temporal information regarding tracer kinetics. By capturing the time-varying distribution of radiotracers, dynamic PET is particularly useful for lesion detection, treatment monitoring, and progression assessment \cite{article, 9201038}.  From dynamic PET data, parametric images can be estimated using voxel-level kinetic modeling, providing quantitative maps of underlying physiological processes. Parametric imaging has been extensively utilized across clinical and research domains, such as assessing tumor metabolism, evaluating myocardial perfusion and viability, and analyzing cerebral blood flow \cite{Tomasi2012, Schaefferkoetter2017, Nakajo2021, Ewing2003, meltzer2000does,lu20123}. Because these applications depend on accurate quantitative metrics, obtaining high-quality PET parametric images is critical for reliable analysis.

One common approach for PET parametric image estimation is to first reconstruct dynamic PET frames using conventional reconstruction algorithms and then estimate kinetic parameters by fitting voxel-level time–activity curves (TACs) to a pre-selected kinetic model. However, parametric images derived from dynamic PET data as described above often exhibit low image quality, primarily due to the inherently ill-posed nature of kinetic modeling and the limited counts resulting from non-continuous data acquisition across multiple bed positions. To address these challenges, various methods have been proposed to enhance PET parametric image quality. Denoising filters are frequently applied either to the dynamic PET frames or to the parametric images generated \cite{Turkheimer2008, Dutta2013, Huang2019,Christian2010HYPR}. For example, the nonlocal mean filtering was introduced to denoise dynamic PET images to generate better parametric images \cite {Dutta2013}. Highly Constrained Back Projection (HYPR) denoising \cite{Christian2010HYPR} was proposed and widely utilized for dynamic PET image denoising by creating a composite image from all dynamic PET frames.

Recently, diffusion models \cite{ho2020denoisingdiffusionprobabilisticmodels, song2019, song2020} have emerged as powerful generative frameworks that can transform a simple pre-defined distribution into complex data distributions through iterative stochastic refinements. These models operate by progressively adding noise to images in a forward diffusion process and generating new samples via a reverse denoising process. They have achieved state-of-the-art performance in high-fidelity image generation and restoration for natural images \cite{beatgan, singh2022highfidelityguidedimagesynthesis, DBLP:journals/corr/abs-2201-09865, DBLP:journals/corr/abs-2112-10752, chung2024diffusionposteriorsamplinggeneral}. More recently, diffusion models have also been successfully applied to medical imaging, demonstrating promising results in PET image denoising \cite{Gong2024,xie2023,shen2023pet,Cho2024,pan2024full,Yu2025,webber2025personalized,zhou2024unsupervised,Clementine2025,zhang2024realization}.

In this work, we propose a novel diffusion model–based kinetic modeling framework to enhance PET parametric image quality. Here, we focus on the Patlak model \cite{Patlak83}, a widely used graphical method for tracers with irreversible uptake, including \textsuperscript{18}F-FDG, the most commonly used tracer in oncology. One issue with training a diffusion model for dynamic PET is the limited availability of high-quality dynamic PET datasets, primarily due to the long acquisition times required for dynamic PET imaging. Further increasing the acquisition time or dose levels to obtain high-quality dynamic PET data is not feasible. With the availability of total-body PET scanners in clinics, high-quality static PET images are becoming available. As demonstrated in our previous work, the score function trained on normal-dose total-body static PET datasets is of high quality and can provide valuable information for denoising PET images obtained from different scanners, tracers, and dose levels \cite{Yu2025}. In this work, we propose leveraging the high-quality score function trained on a large cohort of total-body \textsuperscript{18}F-FDG static PET data \cite{Yu2025} as a prior, exploiting the patch-wise similarity between static and parametric images. The integration of the diffusion model-based prior into the kinetic model is inspired by the recent developments of regularization by denoising (RED) \cite{romano2017littleenginecouldregularization} and RED-Diff~\cite{mardani2023variationalperspectivesolvinginverse} frameworks. The half-quadratic splitting (HQS)-based optimization framework is adopted to decouple the kinetic modeling fitting and the diffusion model-guided denoising steps, as these two processes have different requirements for iteration numbers. To further enhance computational efficiency, we incorporate initializations from regular kinetic modeling to reduce the number of required diffusion steps. The proposed framework is validated on two clinical total-body dynamic PET datasets, including COVID-19 \cite{wang2023total} and genitourinary cancer (GUC) cohorts \cite{li2024totalbodyparametricimagingusing}. Comparisons under both normal-dose and low-dose conditions demonstrate the effectiveness of the proposed framework. 

\section{Method}
\subsection{Diffusion Models}
Diffusion models \cite{ho2020denoisingdiffusionprobabilisticmodels} consist of two processes, a forward process that gradually corrupts the data by injecting Gaussian noise, and a reverse process that recovers data from the noise. The forward process can be modeled as a Markov process,
\begin{equation}
\begin{aligned}
    q(\boldsymbol{x}_{1:T}\mid\boldsymbol{x}_0) &:= \prod \limits_{t=1}^T q(\boldsymbol{x}_t\mid \boldsymbol{x}_{t-1}), \\
    q(\boldsymbol{x}_t\mid\boldsymbol{x}_{t-1}) &:= \mathcal{N}(\boldsymbol{x}_t; \sqrt{1-\beta_t}\boldsymbol{x}_{t-1}, \beta_t \boldsymbol{I}),
\label{diff-forward}
\end{aligned}
\end{equation}
where $\beta_t$ denotes the variance of the Gaussian noise added at each time step $t$. One property of this forward process is that we can sample $\boldsymbol{x}_t$ at any arbitrary step $t$ directly conditioned on
$\boldsymbol{x}_0$. Defining $\alpha_t := 1-\beta_t$, $\bar{\alpha}_t := \prod_{s=1}^t \alpha_s$, we can have
\begin{equation}
\begin{aligned}
    q(\boldsymbol{x}_t\mid\boldsymbol{x}_0) &= \mathcal{N}(\boldsymbol{x}_t; \sqrt{\bar{\alpha}_t} \boldsymbol{x}_0, (1 - \bar{\alpha}_t) \boldsymbol{I}), \\
    \boldsymbol{x}_t &= \sqrt{\bar{\alpha}_t} \boldsymbol{x}_0 + \sqrt{1 - \bar{\alpha}_t} \boldsymbol{\epsilon}, \quad \boldsymbol{\epsilon} \sim \mathcal{N}(\boldsymbol{0}, \boldsymbol{I}).
\label{q_xt_x0}
\end{aligned}
\end{equation}
According to Bayes' theorem, the posterior distribution $q(\boldsymbol{x}_{t-1}\mid\boldsymbol{x}_t)$ also follows a Gaussian distribution
\begin{equation}
    q(\boldsymbol{{x}}_{t-1}\mid\boldsymbol{x}_t, \boldsymbol{x}_0) = \mathcal{N}(\boldsymbol{x}_{t-1}; \boldsymbol{\tilde{\mu}}_t(\boldsymbol{x}_t, \boldsymbol{x}_0), \tilde{\beta}_t \boldsymbol{I}),
\label{q_xt-1_xt}
\end{equation}
where the mean $\boldsymbol{\tilde{\mu}}_t(\boldsymbol{x}_t, \boldsymbol{x}_0)$ and variance $\tilde{\beta}_t$ can be calculated as
\begin{equation}
\begin{aligned}
    \boldsymbol{\tilde{\mu}}_t(\boldsymbol{x}_t, \boldsymbol{x}_0) &= \frac{\sqrt{\bar{\alpha}_{t-1}} \beta_t}{1 - \bar{\alpha}_t} \boldsymbol{x}_0 
    + \frac{\sqrt{\alpha_t} (1 - \bar{\alpha}_{t-1})}{1 - \bar{\alpha}_t} \boldsymbol{x}_t, \\
    \tilde{\beta}_t &= \frac{1 - \bar{\alpha}_{t-1}}{1 - \bar{\alpha}_t} \beta_t.   
\end{aligned}
\end{equation}
Ideally, $\boldsymbol{x}_0$ can be generated by first sampling from $q(\boldsymbol{{x}}_{T})$, an isotropic Gaussian distribution, then sampling from the posterior distribution $q(\boldsymbol{{x}}_{t-1}\mid\boldsymbol{x}_t)$. However, as the true distribution of $\boldsymbol{x}_0$ is unknown, directly computing $q(\boldsymbol{x}_{t-1}\mid\boldsymbol{x}_t)$ is infeasible. To address this, the denoising diffusion probabilistic model (DDPM) \cite{ho2020denoisingdiffusionprobabilisticmodels} approximates $q(\boldsymbol{x}_{t-1}\mid\boldsymbol{x}_t)$ using a parameterized neural network $p_\gamma(\boldsymbol{x}_{t-1}\mid\boldsymbol{x}_t)$, defined as
\begin{equation}
    p_\gamma(\boldsymbol{x}_{t-1}\mid\boldsymbol{x}_t) := \mathcal{N}(\boldsymbol{x}_{t-1}; \boldsymbol{\tilde{\mu}}_\gamma(\boldsymbol{x}_t, t), \sigma_t^2 \boldsymbol{I}),
\end{equation}
where $\boldsymbol{\tilde{{\mu}}}_\gamma(\boldsymbol{x}_t, t)$ is the mean and $\sigma_t^2$ is the noise variance. Rather than directly modeling $\boldsymbol{\tilde{{\mu}}}_\gamma(\boldsymbol{x}_t, t)$, Ho et al. \cite{ho2020denoisingdiffusionprobabilisticmodels} propose an alternative formulation that models the noise term $\boldsymbol{\epsilon}$ in (\ref{q_xt_x0}) using a learnable function $\boldsymbol{\epsilon}_\gamma(\boldsymbol{x}_t, t)$. This noise term $\boldsymbol{\epsilon}$ is proportional to the gradient of the data log-likelihood, commonly referred to as the score function. Here we directly refer to $\boldsymbol{\epsilon}_\gamma(\boldsymbol{x}_t, t)$ as the score function for simplicity. Consequently, $\boldsymbol{\tilde{\mu}}_\gamma(\boldsymbol{x}_t, t)$ can be written as:
\begin{equation}
    \boldsymbol{\tilde{\mu}}_\gamma(\boldsymbol{x}_t, t) = \frac{1}{\sqrt{\alpha_t}} \left( \boldsymbol{x}_t - \frac{\beta_t}{\sqrt{1 - \bar{\alpha}_t}} \boldsymbol{\epsilon}_\gamma(\boldsymbol{x}_t, t) \right).
\end{equation}
With the trained score function ${\boldsymbol{\epsilon}}_{\hat{\gamma}}
(\boldsymbol{x}_t, t)$, each refinement step during inference can be written as
\begin{equation}
    \boldsymbol{x}_{t-1} = \frac{1}{\sqrt{\alpha_t}} 
    \left( \boldsymbol{x}_t - \frac{\beta_t}{\sqrt{1 - \bar{\alpha}_t}} 
    \boldsymbol{\epsilon}_{\hat{\gamma}}(\boldsymbol{x}_t, t) \right) + \sigma_t \boldsymbol{z},
\label{eq:ddpm_update}
\end{equation}
where $\boldsymbol{z} \sim \mathcal{N}(0, \boldsymbol{I})$.

\subsection{Patlak Model for Parametric Imaging}
The Patlak graphical model \cite{Patlak83} is suitable for systems characterized by irreversible compartment models. According to the Patlak model, the tracer activity concentration at time $t$, denoted as $\boldsymbol{c}(t; \boldsymbol{x}) \in \mathbb{R}^{J \times 1}$, can be expressed as
\begin{equation}
    \boldsymbol{c}(t; \boldsymbol{x}) = \boldsymbol{\kappa} \int_0^{t} C_p(\tau)  d\tau + \boldsymbol{b} C_p(t), \quad t > t^\ast,
\label{tracer_concentration}
\end{equation}
where $C_p(t)$ denotes the blood input function, $J$ indicates the number of image pixels,  $t^\ast$ represents the time when the tracer reaches a steady state, $\boldsymbol{\kappa} \in \mathbb{R}^{J \times 1}$ is the Patlak slope image, and $\boldsymbol{b} \in \mathbb{R}^{J \times 1}$ corresponds to the Patlak intercept image. The PET image intensity at the $m$-th frame $\boldsymbol{y}_m$ can be written as 
\begin{equation}
\boldsymbol{y}_m = \int_{t_{s,m}}^{t_{e,m}} \boldsymbol{c}(\tau) \, d\tau = \bar{S}_p(m) \boldsymbol{\kappa} + \bar{C}_p(m) \boldsymbol{b},
\label{frame_formula}
\end{equation}
where $t_{s, m}$ and $t_{e, m}$ denote the start and end times of frame $m$, respectively.   $\bar{S}_p(m)$ and $\bar{C}_p(m)$ are calculated as
\begin{equation}
\begin{aligned}
    \bar{S}_p(m) &= \int_{t_{s,m}}^{t_{e,m}} \int_0^\tau C_p(\tau_1) d\tau_1 \, d\tau, \\
    \bar{C}_p(m) &= \int_{t_{s,m}}^{t_{e,m}} C_p(\tau) \, d\tau.
\end{aligned}
\end{equation}
Concatenating $\boldsymbol{\kappa}$ and $\boldsymbol{b}$,  the parametric image vector $\boldsymbol{x}$ is defined as
\begin{equation}
\boldsymbol{x} = \begin{bmatrix}\boldsymbol{\kappa}', \boldsymbol{b}'\end{bmatrix}'.
\end{equation}
Correspondingly, the mean of PET images at all $M$ frames, $\bar{\boldsymbol{y}} = [\bar{\boldsymbol{y}}_1', \bar{\boldsymbol{y}}_2', ..., \bar{\boldsymbol{y}}_M']' $, can be related to the parametric image vector $\boldsymbol{x}$ based on the Kronecker product as 
\begin{equation}
    \bar{\boldsymbol{y}} = (\boldsymbol{B} \otimes \boldsymbol{I}_J) \boldsymbol{x},
\label{patlak_inverse_problem}
\end{equation}
where $\boldsymbol{I}_J$ is a $J \times J$ identity matrix and $\boldsymbol{B} \in \mathbb{R}^{M \times 2}$ represents the Patlak temporal basis matrix as
\begin{equation}
\boldsymbol{B} = \begin{bmatrix}
\bar{S}_p(1) & \bar{C}_p(1) \\
\bar{S}_p(2) & \bar{C}_p(2) \\
\vdots & \vdots \\
\bar{S}_p(M) & \bar{C}_p(M)
\end{bmatrix}.
\end{equation}
As we can tell, the Patlak graphical model provides a linear representation of irreversible tracer kinetics. For simplicity of the following derivations, we define $\boldsymbol{A}$ as $\boldsymbol{A} = \boldsymbol{B} \otimes \boldsymbol{I}_J$ in (\ref{patlak_inverse_problem}), where $\boldsymbol{A} \in \mathbb{R}^{MJ \times 2J}$. Assuming that the noise in dynamic PET images $\boldsymbol{e}$ follows a Gaussian distribution, $\boldsymbol{e} \sim \mathcal{N}(0, \sigma^2 \boldsymbol{I})$, the measurement model can be expressed as 
\begin{equation}
{\boldsymbol{y}} = \boldsymbol{A} \boldsymbol{x} + \boldsymbol{e}. 
\end{equation}
Thus, the corresponding likelihood function is given by 
\begin{equation}
p(\boldsymbol{y}\mid\boldsymbol{x}) = \mathcal{N}(\boldsymbol{y}; \boldsymbol{Ax}, \sigma^2\boldsymbol{I}). 
\label{likelihood}
\end{equation}

\subsection{Patlak Estimation Based on Diffusion Models}
The Patlak parametric images can be estimated directly through least squares-based fitting, either iteratively or analytically. To improve parametric image quality, we propose to incorporate prior information through diffusion models into the estimation process by reformulating the estimation as sampling from the posterior distribution $p(\boldsymbol{x}\mid\boldsymbol{y})$. By Bayes’ theorem, the gradient of the posterior distribution can be expressed as
\begin{equation}
    \nabla_{\boldsymbol{x}} \log p(\boldsymbol{x}\mid\boldsymbol{y})
    = \nabla_{\boldsymbol{x}} \log p(\boldsymbol{y}\mid\boldsymbol{x})
    + \nabla_{\boldsymbol{x}} \log p(\boldsymbol{x}),
    \label{bayes_decomposition}
\end{equation}
where the first term enforces data fidelity with the measured dynamic PET data, and the second term provides prior regularization obtained from a pre-trained score function.

During reverse sampling of the diffusion process, as shown in  (\ref{eq:ddpm_update}), the data-consistency gradient can be embedded at each denoising step to incorporate dynamic PET information. The gradient of the likelihood term can be approximated as \cite{Gong2024}
\begin{equation}
    \nabla_{\boldsymbol{x}_t} \log p(\boldsymbol{y}\mid\boldsymbol{x}_t)
    \simeq
    -\frac{1}{\sigma^2}
    \nabla_{\boldsymbol{x}_t}
    \|\boldsymbol{y} - \boldsymbol{A}\boldsymbol{x}_t\|_2^2.
    \label{eq:gradient_log_likelihood}
\end{equation}
By integrating this data-consistency term into the standard diffusion reverse sampling as shown in equation (\ref{eq:ddpm_update}), the inference step can be formulated as
\begin{equation}
\begin{aligned}
    \boldsymbol{x}_{t-1} &= \frac{1}{\sqrt{\alpha_t}} 
    \Bigg( \boldsymbol{x}_t - \frac{\beta_t}{\sqrt{1 - \bar{\alpha}_t}} 
    \boldsymbol{\epsilon}_{\hat{\gamma}}(\boldsymbol{x}_t, t) \Bigg) \\
    &\quad - \frac{\sigma_t^2}{\sigma^2} 
    \nabla_{\boldsymbol{x}_t} 
    \|\boldsymbol{y} - \boldsymbol{A}\boldsymbol{x}_t\|_2^2 \\
    &\quad + \sigma_t z, \quad \text{where} \quad 
    z \sim \mathcal{N}(\boldsymbol{0}, \boldsymbol{I}).
\end{aligned}
\label{inference}
\end{equation}

\begin{figure}[!t]
    \centering
    \includegraphics[width=\columnwidth]{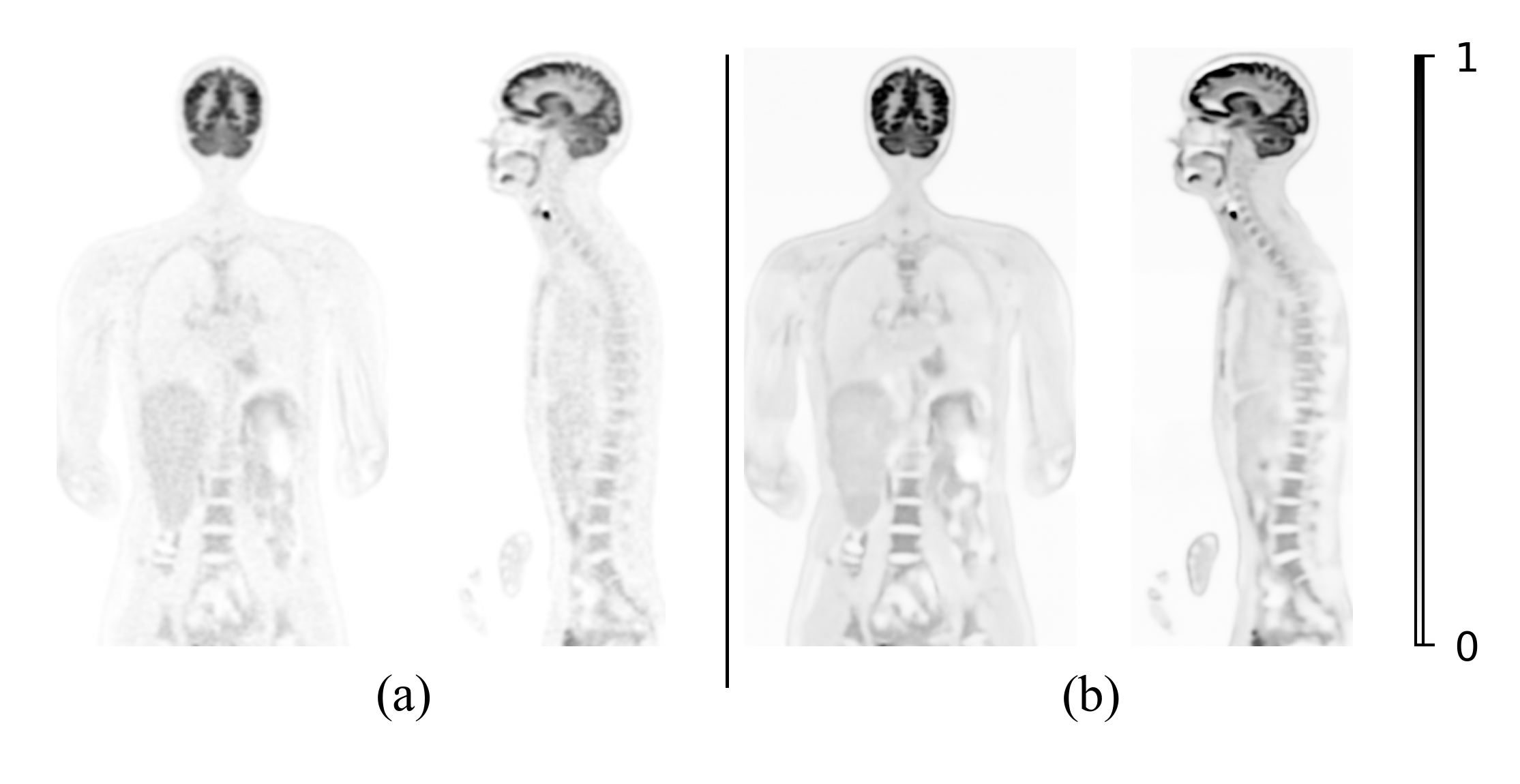}
    \caption{Comparison of estimated Patlak intercept image. (a) Gaussian filtering. (b) DDPM-based method.}
    \label{fig_gaussian_ddpm}
\end{figure}

Obtaining a well-trained score function is challenging due to the limited availability of high-quality dynamic PET datasets. To overcome this limitation, we propose to utilize the score function trained on total-body static PET images. The preliminary results in our conference paper \cite{huang2025patlak} are based on this idea and rely on the inference step as described in equation (\ref{inference}). Fig.~\ref{fig_gaussian_ddpm} compares the coronal and sagittal views of Patlak slope images reconstructed using Gaussian filtering and the proposed inference steps as described in equation (\ref{inference}). Although the diffusion model-based method achieves improved denoising performance and better preservation of structural details, the reconstructed parametric images exhibit elevated overall intensity and nonzero background, indicating incomplete convergence of the kinetic modeling step. This incomplete convergence arises from the coupling between the kinetic modeling step and the reverse denoising procedure during the inference process. 
 
\subsection{Patlak Estimation Based on RED-Diff and HQS}
To further enhance the quality of the reconstructed parametric images and address the convergence issue mentioned above, we propose an improved approach that utilizes diffusion models as a prior term to regularize the Patlak-model fitting, thereby explicitly decoupling kinetic modeling from the diffusion reverse steps. Inspired by RED \cite{romano2017littleenginecouldregularization} and its recent extension to diffusion models \cite{mardani2023variationalperspectivesolvinginverse}, we formulate the optimization objective as a combination of a data fidelity term and a diffusion model-based regularization term,
\begin{equation}
    \hat{\boldsymbol{x}} = \arg\min_{\boldsymbol{x}} \frac{1}{2} \left\| \boldsymbol{y} -  \boldsymbol{A}{\boldsymbol{x}} \right\|_2^2 + \mathcal{R}_{\text{RED}}(\boldsymbol{x}),
\label{red-diff-formula}
\end{equation}
where the regularization term $\mathcal{R}_{\text{RED}}(\boldsymbol{x})$ is defined as
\begin{equation}
    \mathcal{R}_{\text{RED}} = \mathbb{E}_{t \sim \mathcal{U}[0, T], \boldsymbol{\epsilon}_t \sim \mathcal{N}(0, I)} 
    \left[ \omega_t \cdot \mathrm{sg} \left( \boldsymbol{\boldsymbol{\epsilon}}_{\hat{\gamma}}(\boldsymbol{x}_t, t) - \boldsymbol{\epsilon}_t \right)^T \boldsymbol{\mu} \right].
    \label{red_regularizer}
\end{equation}
Here $\omega_t$ serves as a weighting factor. $\boldsymbol{\epsilon}_{\hat{\gamma}}$ represents the noise predicted by the pre-trained score function. The term $\mathrm{sg}(\cdot)$ denotes the stop-gradient operation. The term $\mathrm{sg} \left( \boldsymbol{\boldsymbol{\epsilon}}_{\hat{\gamma}}(\boldsymbol{x}_t, t) - \boldsymbol{\epsilon}_t \right)^T \boldsymbol{\mu}$ represents the projection of the residual of noise onto the image component $\boldsymbol{\mu}$. Ideally, we want the noise vector and the image vector to be orthogonal, i.e., this term should be zero, ensuring that the noise does not contain any information from the image. 

To decouple the kinetic model fitting from the reverse diffusion steps, we employ the HQS algorithm \cite{geman1995nonlinear} to reformulate the objective function in equation (\ref{red-diff-formula}) as 
\begin{equation}
    \hat{\boldsymbol{x}} = \arg\min_{\boldsymbol{x}} \frac{1}{2} \left\| \boldsymbol{y} -  \boldsymbol{A}{\boldsymbol{x}} \right\|_2^2 + \mathcal{R}_{\text{RED}}(\boldsymbol{v}) + \frac{\lambda}{2} \left\| \boldsymbol{x} -  {\boldsymbol{v}} \right\|_2^2,
\label{hqs-objective}
\end{equation}
where $\lambda$ is the penalty parameter.
We solve this by alternatively updating $\boldsymbol{x}$ and $\boldsymbol{v}$, leading to the following subproblems,
\begin{align} 
    {\boldsymbol{x}}^{n+1} &= \arg\min_{\boldsymbol{x}} \frac{1}{2} \left\| \boldsymbol{y} -  \boldsymbol{A}{\boldsymbol{x}} \right\|_2^2 + \frac{\lambda}{2} \left\| \boldsymbol{x} -  {\boldsymbol{v}^n} \right\|_2^2, \label{hqs-subproblems}\\
    \boldsymbol{v}^{n+1} &= \arg\min_{\boldsymbol{v}} \mathcal{R}_{\text{RED}}(\boldsymbol{v}) + \frac{\lambda}{2} \left\| \boldsymbol{x}^{n+1} -  {\boldsymbol{v}} \right\|_2^2. \label{hqs-subproblems2}
\end{align}
For Subproblem (\ref{hqs-subproblems}),  we apply the optimization transfer method to construct a surrogate function $Q_L({\boldsymbol{x\mid x}^k})$ to decouple $\boldsymbol{A}$ and ${\boldsymbol{x}}$ so that each voxel can be optimized independently. The surrogate function for voxel $j$ is
\begin{equation}
\begin{aligned}
    x_{j}^{k+1} &= \arg\min_{x_{j}} \sum\limits_{i=1}^{MJ}\sum\limits_{j=1}^{2J} \frac{1}{2}\alpha_{ij}^k \left(y_{i} - \frac{A_{ij}}{\alpha_{ij}^k} x_{j} \right)^2 \\
    &+ \sum\limits_{j=1}^{2J}\frac{\lambda}{2} \left(x_{j} - v_{j}^k \right)^2,
\label{eq:surrogate-function}
\end{aligned}
\end{equation}
where 
\begin{equation}
\alpha_{ij}^k = \frac{A_{ij} x_{j}^k}{\sum\limits_{j=1}^{2J} A_{ij}x_{j}^k} 
= \frac{A_{ij}x_{j}^k}{A_i \boldsymbol{x}^k}.
\end{equation}
The iterative update equation for $x_{j}^k$ can be expressed as
\begin{equation}
x_{j}^{k, n+1} = x_{j}^{k, n} 
\frac{\sum\limits_{i=1}^{MJ} A_{ij}y_{j} + \lambda v_{j}^k }
{\sum\limits_{i=1}^{MJ} A_{ij}\left[ A_i \boldsymbol{x}^{k, n} \right] + \lambda x_{j}^{k, n}}.
\label{eq:sub-iter-update-x}
\end{equation}
The algorithm to solve Subproblem~(\ref{hqs-subproblems2}) is the same as in \cite{mardani2023variationalperspectivesolvinginverse,bae2025pet}. The algorithm details for the proposed framework are summarized in Algorithm \ref{alg2}.
\begin{algorithm}
\caption{Proposed Patlak image estimation framework.}
\label{alg2}
\textbf{Input:} Pre-trained score function $\boldsymbol{\epsilon}_{\gamma^*}$, Patlak basis matrix $\boldsymbol{A}$, dynamic PET images \( \boldsymbol{y} \), total iterations \(\texttt{MaxIt}\), sub-iteration number \(\texttt{SubIt1}\), sub-iteration number \(\texttt{SubIt2}\)
\begin{algorithmic}[1]
\State Initialize: Run baseline iterative method and apply Gaussian smoothing to obtain \(x^{k, 0}\)
\For{$k = 1$ to $\texttt{MaxIt}$}
    \State $x^{k,0} = v^{k-1}$
    \For{$n = 0$ to $\texttt{SubIt1}$}
        \State $x_{j}^{k, n+1} = x_{j}^{k, n} 
\frac{\sum\limits_{i=1}^{MJ} A_{ij}y_{j} + \lambda v_{j}^k }
{\sum\limits_{i=1}^{MJ} A_{ij}\left[ A_i \boldsymbol{x}^{k, n} \right] + \lambda x_{j}^{k, n}}$
    \EndFor
    \State $x^k = x^{k,\texttt{SubIt1}}$, $v^{k,\texttt{SubIt2}} = x^{k, \texttt{SubIt1}}$
    \For {$t = \texttt{SubIt2}$ to $0$}
        \State $\epsilon_t \sim \mathcal{N}(0, I)$
        \State \(
        \begin{aligned}
            \text{Loss} &\gets \frac{\lambda}{2} \|x^k - v^{k,t}\|_2^2 + \\
            &\omega_t \cdot sg\left( \boldsymbol{\epsilon}_{\gamma^*}\left(v_{\text{scaled}}^{k, t}, t\right) - \epsilon_t\right)^T v^{k, t}
        \end{aligned}
        \)
        \State \text{where} \(v_{\text{scaled}}^{k, t} = \sqrt{\alpha_t} v^{k, t} + \sqrt{1 - \alpha_t} \epsilon_t\)
        \State Run optimization on \text{Loss}, update $v^{k,t-1}$ from $v^{k,t}$
    \EndFor
    \State $v^k = v^{k,0}$
\EndFor
\State \Return $v^\texttt{MaxIt}$
\end{algorithmic}
\end{algorithm}

\subsection{Implementation Details and Reference Methods}
\subsubsection{Pre-trained score function based on static total-body PET datasets}
We employed a public pre-trained score function derived from 3D static total-body PET datasets\cite{Yu2024, Yu2025}. The score function was trained using 317 total-body full-dose $^{18}$F-FDG PET images acquired with the Siemens Biograph Vision Quadra PET/CT scanner, obtained from the Ultra-low Dose PET Imaging Challenge database. Each image had a matrix size of $440 \times 440 \times 640$ with a voxel size of $1.65 \times 1.65 \times 1.65$ mm$^3$. To improve computational efficiency, background regions were cropped, reducing the image size to $192 \times 288 \times 520$. All images were converted to standardized uptake values (SUV). Due to GPU memory limitations, each 3D volume was randomly cropped into $96 \times 96 \times 96$ patches before being used for model training. 

\subsubsection{Proposed Method}
To align with the pre-trained score function, the leg portion and the extraneous background of the dynamic PET images acquired from the EXPLORER total-body scanner were cropped, resulting in an image dimension of $192 \times 288 \times 520$. The voxel size was also rescaled to match the same voxel size used in the score-function training. To optimize sampling efficiency and accommodate GPU memory constraints, each frame was further partitioned into eight patches of size $192 \times 288 \times 136$, with eight axial slices overlapped between adjacent patches. The inference process was conducted on eight NVIDIA A100 GPUs. After inference, the patches were merged using weighted averaging in the overlapping regions to construct the total-body parametric images. The total inference time for each dynamic dataset was approximately 30 minutes.

The baseline iterative method was implemented following equation~(\ref{eq:sub-iter-update-x}), with the penalty parameter set to zero and $\boldsymbol{x}$ initialized as an all-one matrix. The resulting image was smoothed using Gaussian filtering with a full width at half maximum (FWHM) of 3 voxels, and the output was used as the initial estimate $\boldsymbol{x}^0$. This initialization accelerated convergence, as a reduced number of diffusion steps was used instead of performing the full $T$-step reverse process \cite{chung2022comecloserdiffusefasteracceleratingconditionaldiffusion}. The iterative optimization was then performed by alternating between data consistency steps and diffusion-based regularization steps following the proposed framework as outlined in Algorithm \ref{alg2}. The weighting function $\omega_t$ was defined as the inverse of the time-dependent signal-to-noise ratio (SNR), $\omega_t = 1 / {SNR}_t$, where ${SNR}_t := \alpha_t / \sigma_t$. In our implementation, the total number of outer iterations $\texttt{MatIt}$ was set to 20, with 5 inner data-consistency updates per iteration, resulting in 100 total data-consistency iterations. Specifically, the number of time steps in the reverse process $\texttt{SubIt2}$ was set to 10. To ensure convergence, the penalty parameter $\lambda$ was set to 0.2. The optimization was performed using the ADAM optimizer with a learning rate of 0.01.

\begin{figure}[!t]
\centerline{\includegraphics[width=\columnwidth]{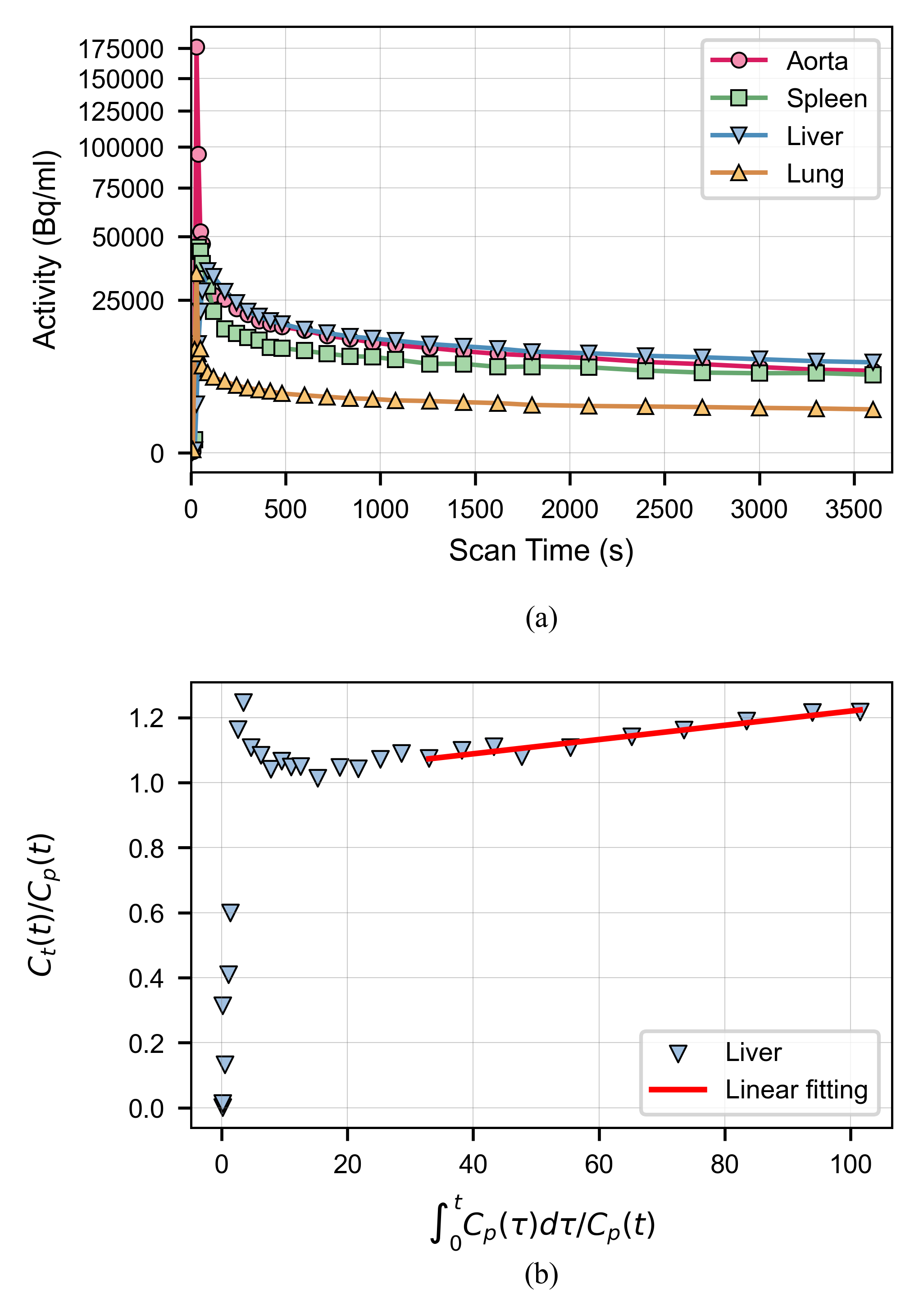}}
\caption{Time activity curves and Patlak fitting of one subject in COVID-19 datasets. (a) Extracted time activity curves of the aorta region and organs. (b) Patlak plot for the liver TAC.}
\label{fig_tac}
\end{figure}
\subsubsection{Reference Methods}
Gaussian filtering and anatomically guided nonlocal means (NLM) filtering \cite{Chan2014, Cui2019PET} were applied to the parametric images reconstructed using the baseline iterative method shown in equation~(\ref{eq:sub-iter-update-x}). In addition, HYPR denoising \cite{Christian2010HYPR} was performed on the full set of dynamic frames, and the denoised data were subsequently reconstructed using the baseline iterative method shown in equation~(\ref{eq:sub-iter-update-x}). These three methods served as comparative baselines to assess the performance of the proposed approach. The Gaussian filter employed a FWHM of 3 voxels. The NLM algorithm was configured with a searching window size of $7 \times 7 \times 7$ voxels. For HYPR denoising, a $3 \times 3 \times 3$ voxel box convolution kernel was used.

\section{Experimental Setup}
\subsection{Datasets}
To further evaluate the proposed framework, experiments were conducted on two clinical total-body dynamic PET datasets. With approval from the Institutional Review Board (IRB) and the Ethics Committee of the University of California, Davis, the study included a cohort of 10 COVID-19 subjects \cite{wang2023total} and 10 Immunotherapy-naïve patients \cite{li2024totalbodyparametricimagingusing} with metastatic genitourinary cancer (GUC). For the COVID-19 cohort, each subject underwent a 1 hour $^{18}$F-FDG total-body dynamic scan on the uEXPLORER PET/CT system following an injection of $^{18}$F-FDG (333 $\pm$ 45 MBq). Similarly, each participant in the GUC cohort received an injection of approximately 370 MBq $^{18}$F-FDG for a 1-hour dynamic PET scan using the uEXPLORER PET/CT system. 

The acquired listmode data were reconstructed into 29 frames with the following timing sequences: $6 \times 10$ s, $2 \times 30$ s, $6 \times 60$ s, $5 \times 120$ s, $4 \times 180$ s, $6 \times 300$ s. Only the last 5 frames were utilized for subsequent analyses, corresponding to a total duration of 25 minutes. Each reconstructed frame had an isotropic voxel size of $4 \times 4 \times 4$ mm$^3$ and an image matrix size of $150 \times 150 \times 486$. Image reconstruction was performed using the ordered subset expectation maximization (OSEM) algorithm with 4 iterations and 20 subsets. Standard corrections were applied to the reconstructed images, including random, scatter, attenuation, dead-time, and decay corrections. The low-dose PET datasets were simulated by shortening the acquisition time across all bed positions to mimic reduced tracer dosage conditions, with image reconstruction performed using the same OSEM settings as in the normal-dose protocol. Both normal-dose and low-dose reconstructions were used for subsequent evaluation and comparison of different methods.

Subsequently, the blood input function was derived directly from the dynamic PET images. A 3D free-hand region of interest (ROI) was placed in the ascending aorta region using a threshold-based approach to extract an image-derived input function. Additionally, ellipsoid ROIs were placed in various organs, as illustrated in Fig.~\ref{fig_tac}(a). The liver region was used as the reference for quantification. The Patlak plot of the liver region along with the fitted line was shown in Fig.~\ref{fig_tac}(b). The observed linearity confirms the physical plausibility of the data and supports the validity of our processing pipeline.

\begin{figure}[!t]
    \centering
    \includegraphics[width=\columnwidth]{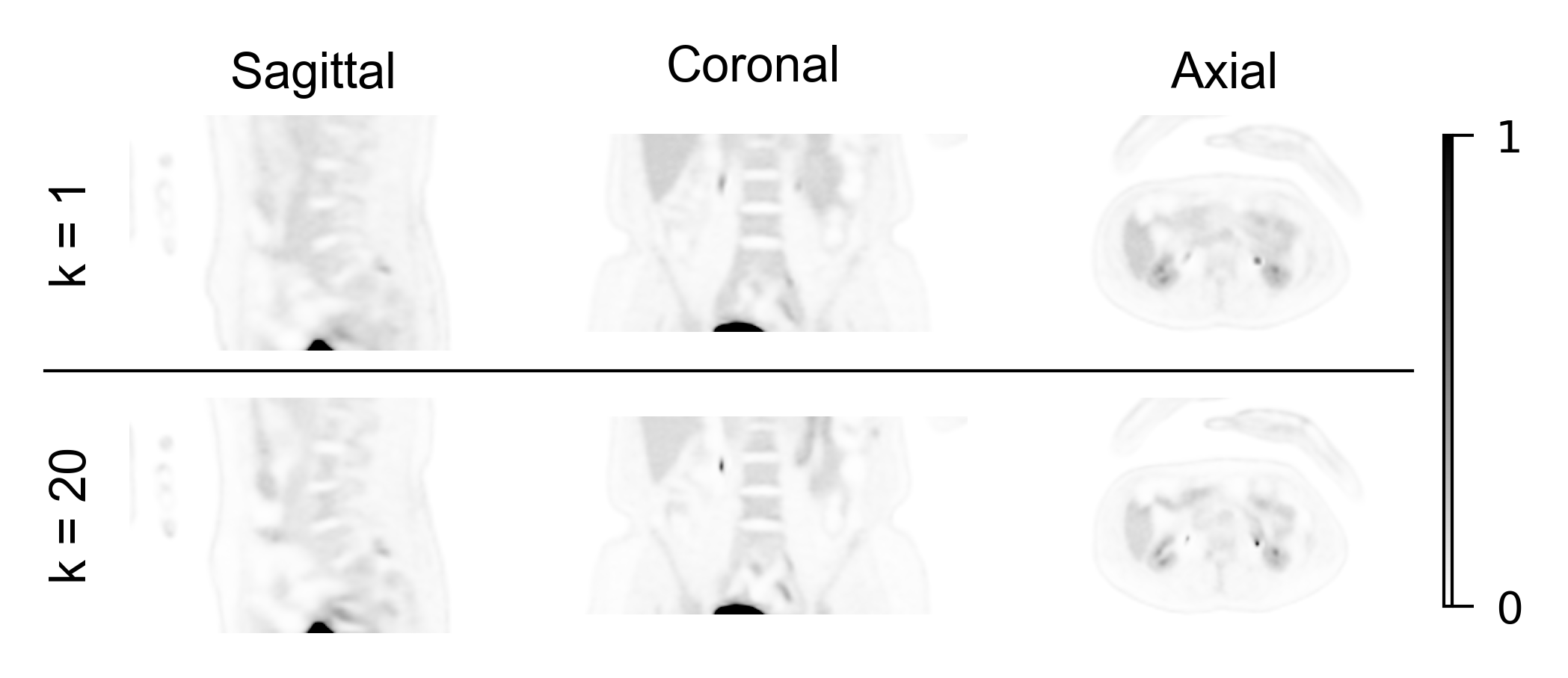}
    \caption{Three-view of one patch showing intermediate results across data consistency iterations for a COVID-19 subject.}
    \label{fig_data_loop}
\end{figure}

\begin{figure}[!t]
    \centering
    \includegraphics[width=\columnwidth]{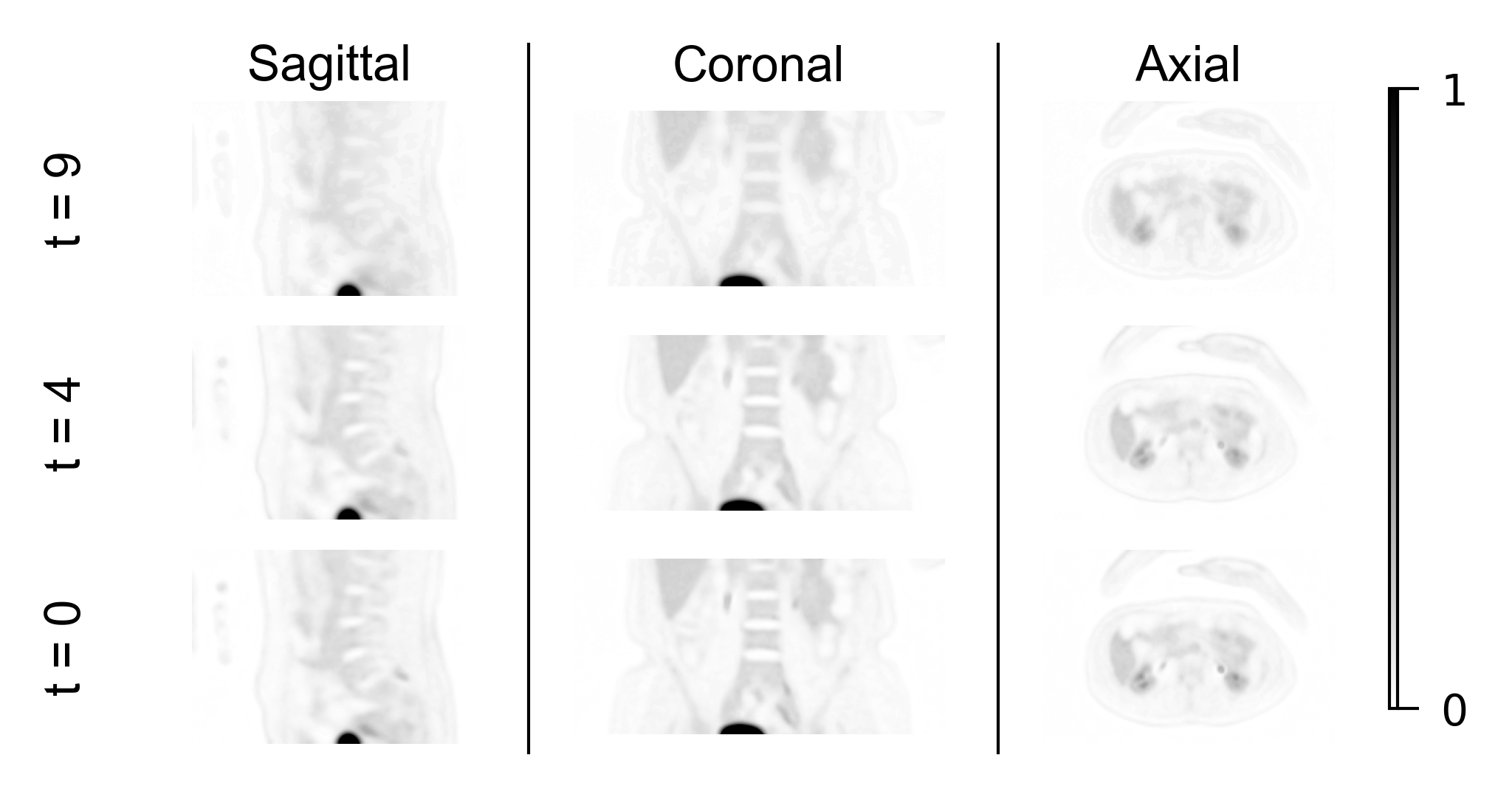}
    \caption{Three-view of the same patch showing intermediate results across denoising iterations for a COVID-19 subject.}
    \label{fig_dn_loop}
\end{figure}

\begin{figure*}[!t]
    \centering
    \includegraphics[width=0.8\textwidth]{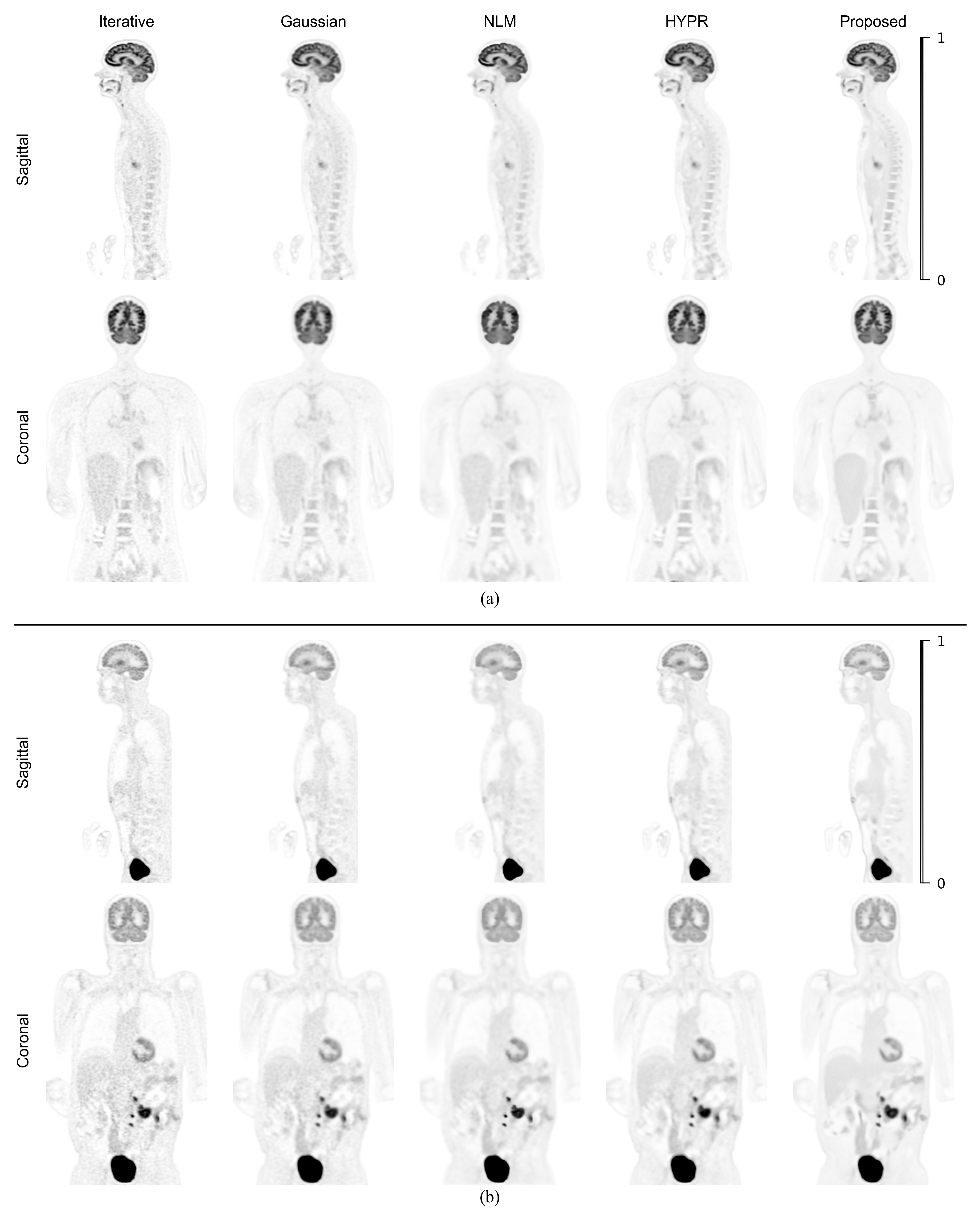}
    \caption{Sagittal and coronal views of Patlak slope images reconstructed from normal-dose datasets using different methods, including the baseline iterative method, Gaussian filtering, NLM, HYPR, and the proposed method. Results are shown for (a) one subject from the COVID-19 dataset and (b) one subject from the GUC dataset.}
    \label{fig_vis_comparison_normal_count}
\end{figure*}

\begin{figure*}[!t]
    \centering
    \includegraphics[width=\textwidth]{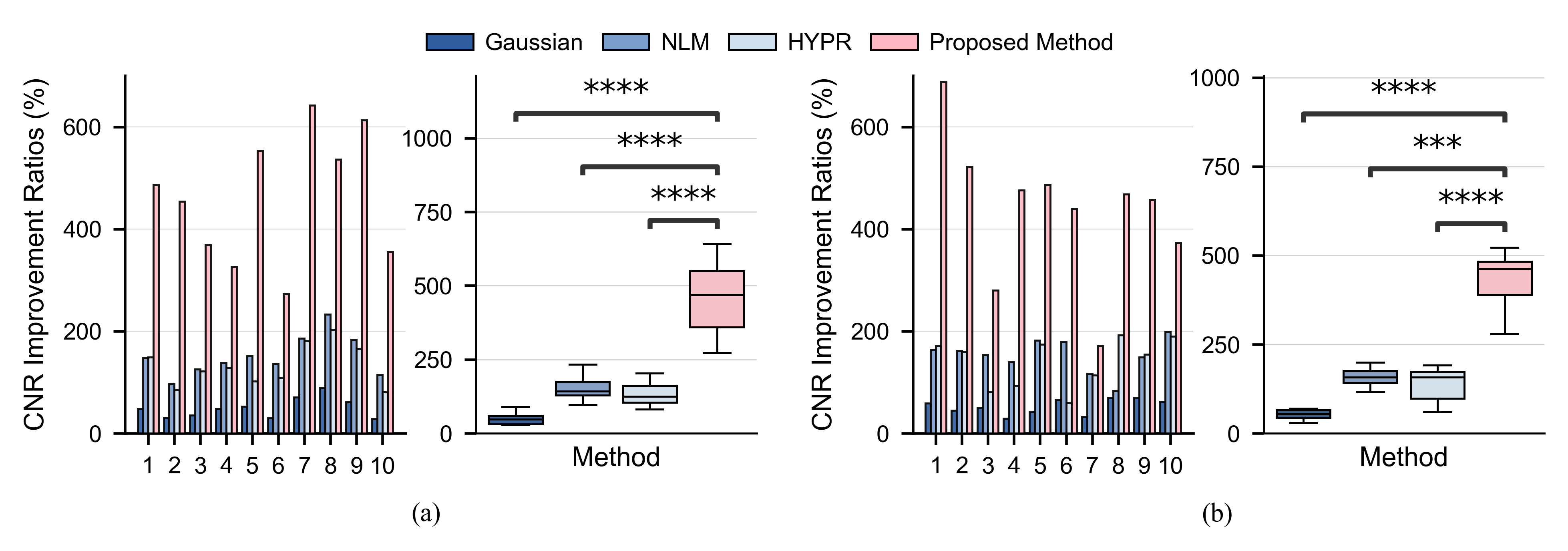}
    \caption{CNR improvement analysis on normal-dose datasets. 
    (a) Patient-wise and summary box plots of CNR improvement ratios for the COVID dataset. 
    (b) Corresponding results for the GUC dataset. 
    Statistical significance levels are denoted as: $p \leq 0.05$ (*), $p \leq 0.01$ (**), $p \leq 0.001$ (***), and $p < 0.00001$ (****).}
    \label{fig_cnr_normal_count}
\end{figure*}

\begin{figure*}[!t]
    \centering
    \includegraphics[width=0.8\textwidth]{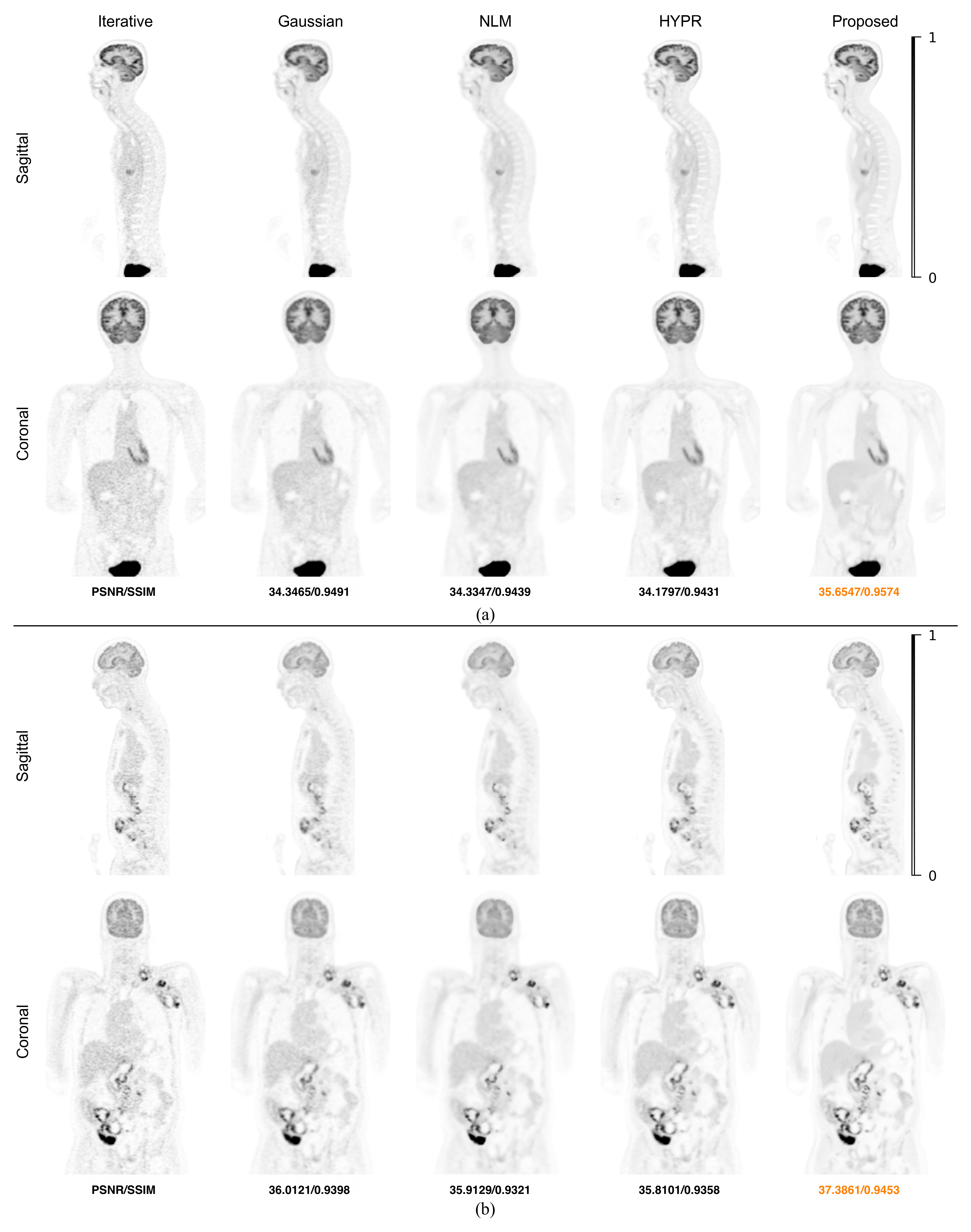}
    \caption{Sagittal and coronal views of Patlak slope images reconstructed from low-dose datasets using different methods. Note that the results for the baseline iterative method (first column) are based on normal-dose data for reference purposes. Results are shown for (a) one subject from the COVID-19 dataset and (b) one subject from the GUC dataset.}
    \label{fig_vis_comparison_low_count}
\end{figure*}

\begin{figure*}[!t]
    \centering
    \includegraphics[width=\textwidth]{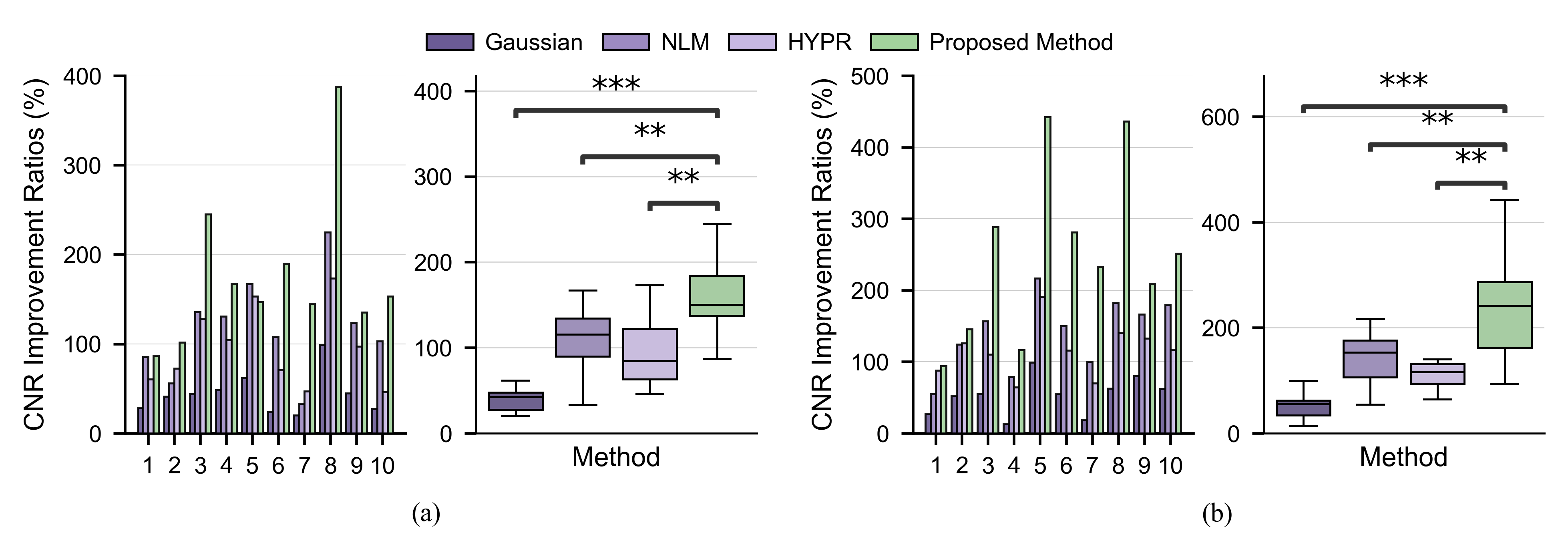}
    \caption{CNR improvement analysis on 1/10 low-dose datasets. 
    (a) Patient-wise and summary box plots of CNR improvement ratios for the COVID dataset. 
    (b) Corresponding analysis for the GUC dataset. 
    Statistical significance levels are denoted as: $p \leq 0.05$ (*), $p \leq 0.01$ (**), $p \leq 0.001$ (***), and $p < 0.00001$ (****).}
    
    \label{fig_cnr_low_count}
\end{figure*}

\begin{figure*}[!t]
    \centering
    \includegraphics[width=\textwidth]{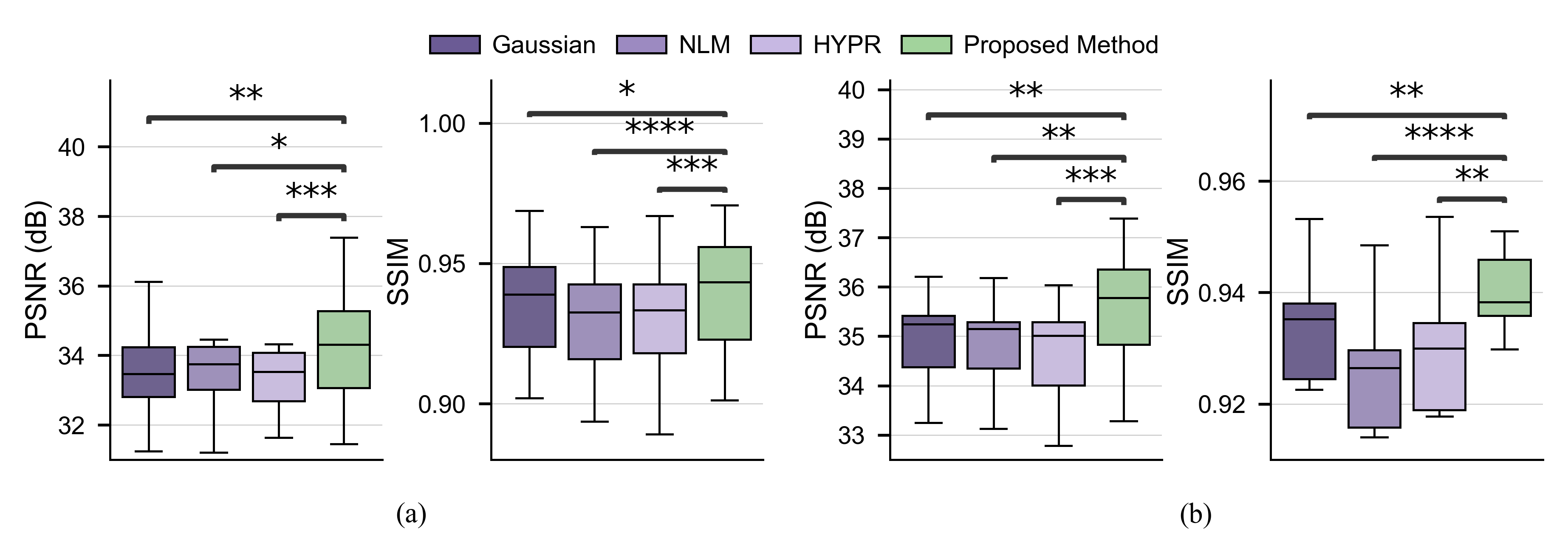}
    \caption{(a) PSNR and SSIM for the COVID dataset across different methods. 
    (b) Corresponding results for the GUC dataset. Statistical significance is shown as $p \leq 0.05$ (*), $p \leq 0.01$ (**), $p \leq 0.001$ (***), $p < 0.00001$ (****), and “ns” for $p \geq 0.05$.}
    
    \label{fig_psnr_ssim}
\end{figure*}
\subsection{Data Analysis} 
To quantitatively evaluate the performance of different methods under varying dose levels, comparisons were conducted using both normal-dose and 1/10 low-dose datasets. Evaluation metrics included peak signal-to-noise ratio (PSNR), structural similarity index measure (SSIM), and contrast-to-noise ratio (CNR). PSNR assessed reconstruction fidelity by quantifying the intensity difference between the reconstructed and reference images. It was calculated based on the maximum possible image intensity ($M$) and the overall pixel-wise mean squared error (MSE) between the two images, as follows:
\begin{equation}
    \text{PSNR} = 10 \cdot \log_{10} \left( \frac{{M}^2}{\text{MSE}} \right).
\end{equation}
SSIM was also utilized to evaluate the image quality. The normal-dose parametric images obtained from the baseline iterative method were used as the reference for evaluating the low-dose results. CNR between the lesion region and the reference region was computed as follows,
\begin{equation}
    \text{CNR} = \frac{\mu_\text{lesion} - \mu_\text{ref}}{\sigma_\text{ref}}
\end{equation}
where $\mu_\text{lesion}$ and $\mu_\text{ref}$ represented the mean intensity values in the lesion and reference regions, respectively, while $\sigma_\text{lesion}$ denoted the voxel-wise standard deviation in the reference region. A total of 20 spherical ROIs (radius = 5 voxels) were selected from the liver region as background regions for noise estimation. The CNR improvement ratio was calculated for each method across different dose-level datasets, using the CNR from the baseline iterative method-based results as the baseline. To assess statistical significance, an independent t-test was performed to compare the CNR improvement ratios between different methods.

\section{Results}
\subsection{Analysis of Intermediate Reconstruction Results}
To assess the effect of the alternating updates, the intermediate results are presented in Fig.~\ref{fig_data_loop} and Fig.~\ref{fig_dn_loop}. Fig.~\ref{fig_data_loop} shows sagittal, coronal, and axial views of a patch from the intermediate results during the data-consistency sub-iterations. The figure compares the first $(k = 1)$ and last $(k = 20)$ sub-iterations. As the iteration progressed, structural details were gradually recovered, the intensity level gradually approached the reference range, and the image contrast improved. These results indicate that the data-consistency updates aligned the reconstructed intensity with the reference range and enhanced recovery of structural details. Fig.~\ref{fig_dn_loop} shows the same patch during the denoising sub-iterations. As the diffusion time step decreased, the noise was progressively reduced while structural details were preserved. The changes in intensity and noise patterns across iterations demonstrate the individual contributions of the data-consistency and diffusion-model parts within the proposed framework.

\subsection{Evaluation on Normal-Dose Datasets}
We first evaluated the performance of the proposed method on normal-dose total-body datasets. Fig.~\ref{fig_vis_comparison_normal_count} shows two views of the estimated Patlak slope images reconstructed using different methods for one subject from the COVID-19 dataset and one subject from the GUC dataset. The reconstructed slope image using the baseline iterative method exhibited high noise. Although Gaussian, NLM, and HYPR filtering produced smoother images, residual noise remained. In contrast, the proposed method suppressed noise more effectively while maintaining lesion contrast and preserving structural details. Fig.~\ref{fig_cnr_normal_count} shows the CNR values and improvement ratios for the COVID-19 and GUC datasets. Gaussian, NLM filtering, and HYPR provided moderate CNR gains compared with the baseline method, whereas the proposed method achieved consistently higher CNR ratios across all datasets. 

\subsection{Evaluation on Low-Dose Datasets}
Additional experiments using low-dose total-body datasets were conducted to further assess the robustness of the proposed method at varying dose levels. Fig.~\ref{fig_vis_comparison_low_count} presents two views of exemplar cases from the COVID-19 and GUC datasets. While Gaussian, NLM filtering, and HYPR approaches partially reduced noise, they also caused oversmoothing that reduced image contrast, whereas the proposed method preserved lesion contrast and structural details. Quantitative results summarized in Fig.~\ref{fig_cnr_low_count} show that the proposed method achieved the highest CNR improvement ratios across all subjects. To further evaluate the image quality, PSNR and SSIM metrics were calculated for the results from the low-dose datasets. Consistent with the CNR evaluation, the results in Fig.~\ref{fig_psnr_ssim} show that the proposed method achieved the best PSNR and SSIM performance.

\section{Discussion}
In this study, we developed a diffusion model-based framework for Patlak parametric image estimation using RED-diff and HQS-based optimization. The method can improve the quality of estimated Patlak images across different dose levels and does not present the convergence issue. The method does not rely on high-quality dynamic PET images as training data, which are difficult to obtain due to acquisition-time and dose constraints. Instead, the score function is pre-trained on static PET images, and the dynamic data and kinetic modeling are used as data constraints. Our experimental results demonstrate that the proposed method is a viable solution for enhancing parametric image estimation.

Our previous implementation \cite{huang2025patlak} directly applied data constraints at each denoising step, resulting in slow convergence due to strong coupling between the kinetic modeling and the diffusion-model sampling. In the present study, the kinetic modeling and the diffusion model are decoupled using the RED-diff framework and HQS-based optimization. This decoupling formulation allows the data-consistency and diffusion-model updates to be optimized alternately, allowing the model to better balance data fidelity and prior information.

The current framework requires manual tuning of several parameters, including the total number of iterations ($\texttt{MaxIt}$), the number of sub-iterations for data consistency ($\texttt{SubIt1}$), the number of diffusion steps ($\texttt{SubIt2}$), and the penalty parameter ($\lambda$). These parameters were empirically adjusted to balance image quality and computational cost. Further investigations on how to pick optimum parameters are needed. Apart from this implementation aspect, the current study focuses on the Patlak model for the FDG tracer. Further extending and evaluating the proposed framework to other kinetic models is needed to test the generalizability of the proposed framework. Finally, apart from total-body datasets, our future work will also further evaluate the proposed framework on dynamic PET datasets from regular PET scanners. 

\section{Conclusion}
In this work, we proposed a diffusion model-based framework for Patlak parametric estimation, leveraging a score function trained on static total-body PET data. The integration of the diffusion prior and kinetic model was achieved following the RED-diff framework and HQS-based optimization. Evaluation on total-body dynamic PET datasets demonstrates that the proposed framework can effectively suppress noise while preserving structural details. Our future work will focus on extending this framework to other kinetic models and evaluating its performance across tracer and scanner types.

\bibliographystyle{IEEEtran}
\bibliography{report}

\end{document}